\documentclass[11pt]{article}
\usepackage{amssymb,amsmath,amsfonts}
\usepackage{graphicx}
\usepackage{graphics}
\usepackage{eepic,epsfig}

=1.60cm

\begin{document}

\title{Casimir effect for curved boundaries \\
in Robertson-Walker spacetime}
\author{A. A. Saharian$^{1}$\thanks{%
E-mail: saharian@ysu.am}\thinspace\ and M. R. Setare$^{2,3}$ \thanks{%
E-mail: rezakord@ipm.ir} \\
\\
\textit{$^1$Department of Physics, Yerevan State University,}\\
\textit{1 Alex Manoogian Street, 0025 Yerevan, Armenia}\vspace{0.3cm}\\
\textit{$^2$Department of Science, Payame Noor University, Bijar, Iran}\vspace{0.3cm} \\
\textit{$^3$Research Institute for Astronomy and Astrophysics of Maragha,}\\
\textit{P.O. Box 55134-441, Maragha, Iran}\\
} \maketitle

\begin{abstract}
Vacuum expectation values of the energy-momentum tensor and the Casimir
forces are evaluated for scalar and electromagnetic fields in the geometry
of two curved boundaries on background of the Robertson-Walker spacetime
with negative spatial curvature. Robin boundary conditions are imposed in
the case of the scalar field and perfect conductor boundary conditions are
assumed for the electromagnetic field. We use the conformal relation between
the Robertson-Walker and Rindler spacetimes and the corresponding results
for two parallel plates moving with uniform proper acceleration through the
Fulling-Rindler vacuum. For the general scale factor the vacuum
energy-momentum tensor is decomposed into the boundary free and boundary
induced parts. The latter is non-diagonal. The Casimir forces are directed
along the normals to the boundaries. For Dirichlet and Neumann scalars and
for the electromagnetic field these forces are attractive for all
separations.
\end{abstract}

\bigskip


\bigskip

\section{Introduction}

An interesting topic in the investigations of the Casimir effect
(for a review see \cite{Eliz94}) is the dependence of the vacuum
properties on the geometry of background spacetime. Analytic
solutions can usually be found only for highly symmetric bulk and
boundary geometries. In particular, motivated by the braneworld
scenarios, the investigations of the Casimir effect in anti-de
Sitter (AdS) spacetime have attracted a great deal of attention.
The Casimir effect provides a natural mechanism for stabilizing
the radion field in these models, as required for a complete
solution of the hierarchy problem. In addition, the Casimir energy
gives a contribution to both the brane and bulk cosmological
constants and, hence, has to be taken into account in the
self-consistent formulation of the braneworld dynamics. The
Casimir energy and corresponding Casimir forces for two parallel
branes in AdS spacetime have been investigated in Refs.
\cite{Gold00} and the local Casimir densities were considered in
Refs. \cite{Knap04}. The Casimir effect in higher dimensional
generalizations of AdS spacetime with compact internal spaces has
been considered in \cite{Flac03}.

Another popular background in gravitational physics is de Sitter (dS)
spacetime. Previously the Casimir effect on this background, described in
planar coordinates, is investigated in Refs. \cite{Seta01} for a conformally
coupled massless scalar field. In this case the problem is conformally
related to the corresponding problem in Minkowski spacetime and the vacuum
characteristics are generated from those for the Minkowski counterpart
multiplying by the conformal factor. The Casimir densities for a massive
scalar field with an arbitrary curvature coupling parameter are considered
in \cite{Saha09}. In \cite{Saha04} the vacuum expectation value of the
energy-momentum tensor for a conformally coupled scalar field is
investigated in dS spacetime with static coordinates in presence of curved
branes on which the field obeys the Robin boundary conditions with
coordinate dependent coefficients. In these papers the conformal relation
between dS and Rindler spacetimes and the results for the Rindler
counterpart were used.

Continuing our previous work \cite{Saha10}, in this paper we study an
exactly solvable problem with bulk and boundary polarizations of the vacuum
on background of less symmetric Robertson-Walker spacetime with negative
spatial curvature. The vacuum expectation values of the energy-momentum
tensor and the Casimir forces are considered for both scalar and
electromagnetic fields in the geometry of two curved boundaries. The
corresponding vacuum densities are obtained by making use of the relation
between the vacuum expectation values in conformally related problems (see,
for instance, \cite{Birrell}) and the results for two infinite plane
boundaries moving with uniform proper accelerations through the
Fulling-Rindler vacuum \cite{Avag02,Saha06}. A closely related problem for
the evaluation of the energy-momentum tensor of a Casimir apparatus in a
weak gravitational field recently has been considered in \cite{Full07,Bimo08}
(see also \cite{Milt10}). In these papers it has been shown that the Casimir
energy for a configuration of parallel plates gravitates according to the
equivalence principle.

The organization of the present paper is as follows. In section \ref%
{sec:Scalar} the bulk and boundary geometries are specified for the problem
under consideration. The vacuum expectation value of the energy-momentum
tensor and the Casimir forces are investigated for a scalar field in the
geometry of two curved boundaries with Robin boundary conditions. The case
of the electromagnetic field with perfect conductor boundary conditions on
the boundaries is discussed in section \ref{sec:Elmag}. The main results are
summarized in section \ref{sec:Conc}.

\section{Scalar field}

\label{sec:Scalar}

\begin{figure}[tbph]
\begin{center}
\epsfig{figure=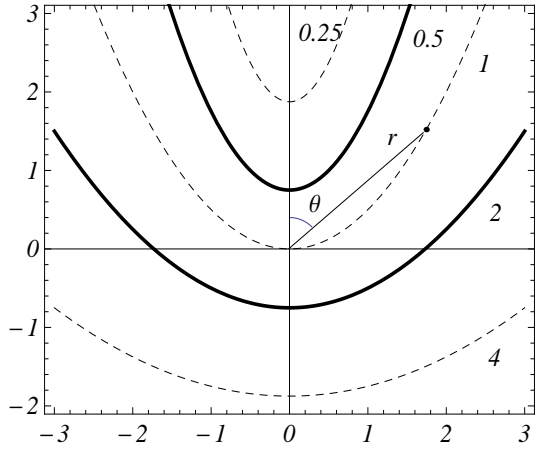,width=7.cm,height=6.cm}
\end{center}
\caption{Thick lines correspond to boundaries (\protect\ref{BoundEq}) with $%
c_{a}=2$ and $c_{b}=0.5$ in the $(r,\protect\theta )$ plane. Dashed curves
are the lines with constant $\protect\xi $. The numbers near the curves are
the corresponding values of $\protect\xi _{0}/\protect\xi $.}
\label{fig1}
\end{figure}

We consider a conformally coupled massless scalar field $\varphi (x)$ on
background of $(D+1)$-dimensional Robertson-Walker (RW) spacetime with
negative spatial curvature. In the hyperspherical coordinates $(r,\theta
_{1}=\theta ,\theta _{2},\ldots ,\theta _{D-1})$, the corresponding line
element has the form
\begin{equation}
ds^{2}=g_{ik}dx^{i}dx^{k}=a^{2}(\eta )(d\eta ^{2}-\gamma
^{2}dr^{2}-r^{2}d\Omega _{D-1}^{2}),  \label{dsRW}
\end{equation}%
where $\gamma =1/\sqrt{1+r^{2}}$ and $d\Omega _{D-1}^{2}=d\theta
^{2}+\sum_{k=2}^{D-1}\left( \prod_{l=1}^{k-1}\sin ^{2}\theta _{l}\right)
d\theta _{k}^{2}$ is the line element on the $(D-1)$-dimensional unit sphere
in Euclidean space. In the case of conformal coupling the field equation
takes the form
\begin{equation}
\left( \nabla _{l}\nabla ^{l}+\frac{D-1}{4D}R\right) \varphi (x)=0,
\label{fieldeq}
\end{equation}%
where $R$ is the Ricci scalar for the RW spacetime. We assume that the field
obeys Robin conditions
\begin{equation}
(\beta _{j}+n_{j}^{l}\nabla _{l})\varphi (x)=0,\;x\in S_{j},\;j=a,b,
\label{BC}
\end{equation}%
on two boundaries described by the equations%
\begin{equation}
\sqrt{1+r^{2}}-r\cos \theta =c_{j},  \label{BoundEq}
\end{equation}%
with positive constants $c_{a}$ and $c_{b}$, $c_{a}>c_{b}$. For the region
between the boundaries, the corresponding normal has the components%
\begin{equation}
n_{j}^{l}=\delta _{j}[c_{j}ra(\eta )]^{-1}(0,\gamma ^{-2}(\gamma
-c_{j}),-\sin \theta ,0,\ldots ,0),  \label{nl}
\end{equation}%
with $\delta _{a}=1$ and $\delta _{b}=-1$. The coefficients $\beta _{j}$ in (%
\ref{BC}) will be specified below. We are interested in the vacuum
expectation value (VEV) of the energy-momentum tensor in the
region between two boundaries. In figure \ref{fig1} we have
plotted these boundaries on the $(r,\theta )$ plane by thick lines
for the values $c_{a}=2$ and $c_{b}=0.5$. Note that for points on
boundary (\ref{BoundEq}) one has $0\leqslant \theta
\leqslant \arccos (\sqrt{1-c_{j}^{2}})$ for $c_{j}\leqslant 1$ and $%
0\leqslant \theta \leqslant \pi $ for $c_{j}>1$.

In order to evaluate the VEV of the energy-momentum tensor, we present the
RW line element in the form conformal to the Rindler metric. This can be
done by making use of the coordinate transformation%
\begin{equation}
x^{i}=(\eta ,r,\theta _{1},\ldots ,\theta _{D-1})\rightarrow x^{\prime
i}=(\eta ,\mathbf{x}^{\prime }),  \label{CoordTrans0}
\end{equation}%
with $\mathbf{x}^{\prime }=(x^{\prime 1},x^{\prime 2},...,x^{\prime D})$,
defined by the relations (for the case $D=3$ see \cite{Birrell})%
\begin{eqnarray}
&&x^{\prime 1}\equiv \xi =\xi _{0}\Omega ,\;x^{\prime 2}=\xi _{0}r\Omega
\cos \theta _{2}\sin \theta ,\ldots ,  \notag \\
&&x^{\prime D-1}=\xi _{0}r\Omega \cos \theta _{D-1}\prod_{l=1}^{D-2}\sin
\theta _{l},\;x^{\prime D}=\xi _{0}r\Omega \prod_{l=1}^{D-1}\sin \theta _{l}.
\label{CordTrans}
\end{eqnarray}%
Here $\xi _{0}$ is a constant having the dimension of length and
\begin{equation}
\Omega =\gamma /(1-r\gamma \cos \theta ).  \label{Omega}
\end{equation}%
In coordinates $x^{\prime i}$ the RW line element becomes
\begin{equation}
ds^{2}=g_{ik}^{\prime }dx^{\prime i}dx^{\prime k}=a^{2}(\eta )\xi
^{-2}g_{ik}^{\mathrm{R}}dx^{\prime i}dx^{\prime k}.  \label{dsRW1}
\end{equation}%
The latter is manifestly conformal to the metric $g_{ik}^{\mathrm{R}}$ of
the Rindler spacetime:%
\begin{equation}
g_{ik}^{\mathrm{R}}=\text{diag}(\xi ^{2},-1,\cdots ,-1).  \label{gikR}
\end{equation}

From (\ref{CordTrans}) and (\ref{Omega}) it follows that in coordinates $%
x^{\prime i}$ the boundaries (\ref{BoundEq}) are described by
simple equations $\xi =\xi _{j}\equiv \xi _{0}/c_{j}$. In figure
\ref{fig1} we have plotted the coordinate lines $\xi
=\mathrm{const}$ in the plane $(r,\theta )$ for several values of
the const (numbers near the curves). The thick lines correspond to
the boundaries. The boundaries divide the space into three
regions. In the regions $\xi <\xi _{a}$ and $\xi >\xi _{b}$ the
VEVs are the
same as in the geometry with a single boundary and were investigated in \cite%
{Saha10}. In what follows we shall be concerned with the region between the
boundaries, $\xi _{a}<\xi <\xi _{b}$. Using the standard transformation
formula for the VEVs of the energy--momentum tensor in conformally related
problems, we can generate the results for the RW spacetime from the
corresponding results for a scalar field $\varphi _{\mathrm{R}}$ in the
Rindler spacetime for two infinite plates \cite{Avag02,Saha06} on which the
field obeys the boundary conditions
\begin{equation}
\left( \beta _{\mathrm{R}j}+n^{\prime }{}_{\mathrm{R}j}^{l}\nabla
_{l}^{\prime }\right) \varphi _{\mathrm{R}}(x^{\prime })=0,\quad \xi =\xi
_{j},\quad n^{\prime }{}_{\mathrm{R}j}^{l}=\delta _{j}\delta _{1}^{l},
\label{boundRind}
\end{equation}%
with a constant coefficients $\beta _{\mathrm{R}j}$. Note that in the
Rindler problem $\xi _{j}^{-1}$ is the proper acceleration of the plate. The
corresponding VEV\ of the energy-momentum tensor we shall denote by $\langle
0_{\mathrm{R}}|T_{i}^{k}\left[ g_{lm}^{\mathrm{R}},\varphi _{\mathrm{R}}%
\right] |0_{\mathrm{R}}\rangle $.

In order to find the relation between the coefficients in boundary
conditions (\ref{BoundEq}) and (\ref{boundRind}), we note that under the
conformal transformation $g_{ik}^{\prime }=[a(\eta )/\xi ]^{2}g_{ik}^{%
\mathrm{R}}$, the field $\varphi _{\mathrm{R}}$ transforms in accordance
with
\begin{equation}
\varphi (x^{\prime })=[\xi /a(\eta )]^{(D-1)/2}\varphi _{\mathrm{R}%
}(x^{\prime }).  \label{FieldTrans}
\end{equation}%
Now, comparing boundary conditions (\ref{BC}), (\ref{boundRind}) and taking
into account Eq. (\ref{FieldTrans}), we find:
\begin{equation}
a(\eta )\beta _{j}=\xi _{j}\beta _{\mathrm{R}j}+\delta _{j}(1-D)/2.
\label{relcoef}
\end{equation}%
Note that with these Robin coefficients the boundary conditions
(\ref{BC}) are time independent. From (\ref{relcoef}) it follows
that Dirichlet boundary condition in the problem on the RW bulk
corresponds to Dirichlet condition in the Rindler counterpart. For
the case of Neumann boundary condition in the RW bulk the
corresponding problem in the Rindler spacetime is of the Robin
type with $\xi _{j}\beta _{\mathrm{R}j}=\delta _{j}(D-1)/2$.

The VEV of the energy-momentum tensor in coordinates $(\eta ,\mathbf{x}%
^{\prime })$ is found by using the transformation formula for conformally
related problems (see, e.g., \cite{Birrell}):
\begin{equation}
\langle 0_{\mathrm{RW}}|T_{i}^{k}\left[ g_{lm}^{\prime },\varphi \right] |0_{%
\mathrm{RW}}\rangle =[\xi /a(\eta )]^{D+1}\langle 0_{\mathrm{R}}|T_{i}^{k}%
\left[ g_{lm}^{\mathrm{R}},\varphi _{\mathrm{R}}\right] |0_{\mathrm{R}%
}\rangle +\langle T_{i}^{k}\left[ g_{lm}^{\prime },\varphi \right] \rangle ^{%
\mathrm{(an)}},  \label{conftransemt}
\end{equation}%
where the second term on the right-hand side is determined by the trace
anomaly. In odd spacetime dimensions this term is absent. For a scalar field
$\varphi _{\mathrm{R}}(x^{\prime })$ with boundary conditions (\ref%
{boundRind}), the VEV in the corresponding Rindler problem is presented in
the form:
\begin{equation}
\langle 0_{\mathrm{R}}|T_{i}^{k}[g_{lm}^{\mathrm{R}},\varphi _{\mathrm{R}%
}]|0_{\mathrm{R}}\rangle =\langle \tilde{0}_{\mathrm{R}}|T_{i}^{k}[g_{lm}^{%
\mathrm{R}},\varphi _{\mathrm{R}}]|\tilde{0}_{\mathrm{R}}\rangle +\langle T_{%
\mathrm{(R)}i}^{k}\rangle ^{\mathrm{(b)}},  \label{TikR}
\end{equation}%
where $|\tilde{0}_{\mathrm{R}}\rangle $ is the vacuum state for the Rindler
spacetime in the absence of boundaries. The term \cite{Saha06} \
\begin{eqnarray}
\langle T_{\mathrm{(R)}i}^{k}\rangle ^{\mathrm{(b)}} &=&\langle T_{\mathrm{%
(R)}i}^{k}\rangle ^{(j)}-\frac{2^{2-D}\pi ^{-(D+1)/2}}{\Gamma ((D-1)/2)}%
\delta _{i}^{k}\int_{0}^{\infty }d\lambda \,\lambda ^{D}\int_{0}^{\infty
}d\omega \,  \notag \\
&&\times \Omega _{j\omega }(\lambda a,\lambda b)F^{(i)}[\bar{I}_{\omega
}^{(j)}(\lambda \xi _{j})K_{\omega }(\lambda \xi )-\bar{K}_{\omega
}^{(j)}(\lambda \xi _{j})I_{\omega }(\lambda \xi )],  \label{Tikb}
\end{eqnarray}%
represents the correction due to the presence of the boundaries. In this
formula, $I_{\omega }(z)$ and $K_{\omega }(z)$ are the modified Bessel
functions and for a given function $f(z)$ the barred notations stand for
\begin{equation}
\bar{f}^{(j)}(z)=\delta _{j}\xi _{j}\beta _{\mathrm{R}j}f(z)+zf^{\prime }(z),
\label{barnot}
\end{equation}%
with $j=a,b$. Other notations are as follows:%
\begin{eqnarray}
\Omega _{a\omega }(u,v) &=&\frac{\bar{K}_{\omega }^{(b)}(v)/\bar{K}_{\omega
}^{(a)}(u)}{\bar{I}_{\omega }^{(b)}(v)\bar{K}_{\omega }^{(a)}(u)-\bar{K}%
_{\omega }^{(b)}(v)\bar{I}_{\omega }^{(a)}(u)},\;  \label{Oma} \\
\Omega _{b\omega }(u,v) &=&\frac{\bar{I}_{\omega }^{(a)}(u)/\bar{I}_{\omega
}^{(b)}(v)}{\bar{I}_{\omega }^{(b)}(v)\bar{K}_{\omega }^{(a)}(u)-\bar{K}%
_{\omega }^{(b)}(v)\bar{I}_{\omega }^{(a)}(u)},  \label{Omb}
\end{eqnarray}%
For a given function $g(z)$, the functions $F^{(i)}[g(z)]$ have the form
\begin{eqnarray}
F^{(1)}[g(z)] &=&-Dg^{\prime 2}(z)-\frac{D-1}{z}g(z)g^{\prime }(z)+D\left( 1+%
\frac{\omega ^{2}}{z^{2}}\right) g^{2}(z),  \notag \\
F^{(i)}[g(z)] &=&g^{\prime 2}(z)+\left( \frac{\omega ^{2}}{z^{2}}-\frac{D+1}{%
D-1}\right) g^{2}(z),\quad i=2,\ldots ,D.  \label{Figz}
\end{eqnarray}%
and $F^{(0)}[g(z)]=-F^{(1)}[g(z)]-\sum_{i=2}^{D}F^{(i)}[g(z)]$. In (\ref%
{Tikb}), the term $\langle T_{\mathrm{(R)}i}^{k}\rangle ^{(j)}$ is the part
induced by a single plate at $\xi =\xi _{j}$ when the second plate is absent
and the second term on the right-hand side is the correction due to the
second boundary.

Formula (\ref{Tikb}) with $j=a$ and $j=b$ provides two equivalent
representations for the VEV. Note that the surface divergences on the
boundary at $\xi =\xi _{j}$ are contained in the first term on the
right-hand side of this formula and the second term is finite. The single
boundary induced part in the region $\xi >\xi _{j}$ is given by the formula
\cite{CandD,Saha02}
\begin{equation}
\langle T_{\mathrm{(R)}i}^{k}\rangle ^{(j)}=\frac{-2^{1-D}\delta _{i}^{k}}{%
\pi ^{(D+1)/2}D\Gamma \left( \frac{D-1}{2}\right) }\int_{0}^{\infty
}d\lambda \,\lambda ^{D}\int_{0}^{\infty }d\omega \frac{\bar{I}_{\omega
}^{(j)}(\lambda \xi _{j})}{\bar{K}_{\omega }^{(j)}(\lambda \xi _{j})}%
F^{(i)}[K_{\omega }(\lambda \xi )].  \label{TikRb}
\end{equation}%
The corresponding expression in the region $\xi <\xi _{j}$ is obtained from (%
\ref{TikRb}) by the replacements $I_{\omega }\rightleftarrows K_{\omega }$.

For RW background, similar to (\ref{TikR}), the VEV of the
energy-momentum tensor in coordinates $x^{\prime i}$ is presented
in the form:
\begin{equation}
\langle 0_{\mathrm{RW}}|T_{i}^{k}\left[ g_{lm}^{\prime },\varphi \right] |0_{%
\mathrm{RW}}\rangle =\langle \tilde{0}_{\mathrm{RW}}|T_{i}^{k}\left[
g_{lm}^{\prime },\varphi \right] |\tilde{0}_{\mathrm{RW}}\rangle +\langle
T_{i}^{k}\left[ g_{lm}^{\prime },\varphi \right] \rangle ^{\mathrm{(b)}},
\label{TikdS}
\end{equation}%
where $\langle \tilde{0}_{\mathrm{RW}}|T_{i}^{k}\left[
g_{lm}^{\prime },\varphi \right] |\tilde{0}_{\mathrm{RW}}\rangle $
is the vacuum expectation value in the boundary free RW spacetime
and the part $\langle T_{i}^{k}\left[ g_{lm}^{\prime },\varphi
\right] \rangle ^{\mathrm{(b)}}$ is the correction due to
boundaries (\ref{BoundEq}). For the separate terms on the
right-hand side of (\ref{TikdS}) we have
\begin{eqnarray}
\langle \tilde{0}_{\mathrm{RW}}|T_{i}^{k}\left[ g_{lm}^{\prime },\varphi %
\right] |\tilde{0}_{\mathrm{RW}}\rangle &=&[\xi /a(\eta )]^{D+1}\langle
\tilde{0}_{\mathrm{R}}|T_{i}^{k}[g_{lm}^{\mathrm{R}},\varphi _{\mathrm{R}}]|%
\tilde{0}_{\mathrm{R}}\rangle +\langle T_{i}^{k}\left[ g_{lm}^{\prime
},\varphi \right] \rangle ^{\mathrm{(an)}},  \label{TikdS0} \\
\langle T_{i}^{k}\left[ g_{lm}^{\prime },\varphi \right] \rangle ^{\mathrm{%
(b)}} &=&[\xi /a(\eta )]^{D+1}\langle T_{\mathrm{(R)}i}^{k}\rangle ^{\mathrm{%
(b)}}.  \label{TikdSb}
\end{eqnarray}

Now, the VEV of the energy-momentum tensor in coordinates $x^{i}$
is obtained from (\ref{TikdS}) by the coordinate transformation
$x^{\prime i}\rightarrow x^{i}$. As before, this VEV may be
written in the form of the sum of purely RW and boundary parts:
\begin{equation}
\langle 0_{\mathrm{RW}}|T_{i}^{k}\left[ g_{lm},\varphi \right] |0_{\mathrm{RW%
}}\rangle =\langle \tilde{0}_{\mathrm{RW}}|T_{i}^{k}\left[ g_{lm},\varphi %
\right] |\tilde{0}_{\mathrm{RW}}\rangle +\langle T_{i}^{k}\rangle ^{\mathrm{%
(b)}}.  \label{TikdS1}
\end{equation}%
Using the expression for $\langle \tilde{0}_{\mathrm{R}}|T_{i}^{k}[g_{lm}^{%
\mathrm{R}},\varphi _{\mathrm{R}}]|\tilde{0}_{\mathrm{R}}\rangle $ from \cite%
{Saha02}, for the purely RW part one finds (for the vacuum polarization in
RW spacetimes see \cite{Birrell,Grib94,Bord97} and references therein)
\begin{eqnarray}
&& \langle \tilde{0}_{\mathrm{RW}}|T_{i}^{k}\left[ g_{lm},\varphi \right] |%
\tilde{0}_{\mathrm{RW}}\rangle  =\langle T_{i}^{k}\left[ g_{lm},\varphi %
\right] \rangle ^{\mathrm{(an)}}+\frac{2[a(\eta )]^{-D-1}}{(4\pi
)^{D/2}D\Gamma (D/2)}  \notag \\
&& \quad \times \mathrm{diag}\left( -D,1,\ldots ,1\right)
\int_{0}^{\infty }\frac{\omega ^{D}d\omega }{e^{2\pi \omega
}+(-1)^{D}}\prod_{l=1}^{l_{m}}\left[ \left( \frac{D-1-2l}{2\omega
}\right) ^{2}+1\right] ,  \label{Tik0dSst}
\end{eqnarray}%
where $l_{m}=D/2-1$ for even $D>2$ and $l_{m}=(D-1)/2$ for odd $D>1$, and
the value of the product should be taken 1 for $D=1,2$. In particular, in $%
D=3$ for the anomaly part we have%
\begin{equation}
\langle T_{i}^{k}\left[ g_{lm},\varphi \right] \rangle ^{\mathrm{(an)}}=%
\frac{^{(3)}H_{i}^{k}-^{(1)}H_{i}^{k}/6}{2880\pi ^{2}},  \label{Tikan}
\end{equation}%
where the expressions for the tensors $^{(j)}H_{i}^{k}$ are given in \cite%
{Birrell}. Now it can be easily checked that for the static case, $a(\eta )=%
\mathrm{const}$, one has $\langle \tilde{0}_{\mathrm{RW}}|T_{i}^{k}\left[
g_{lm},\varphi \right] |\tilde{0}_{\mathrm{RW}}\rangle =0$.

After the coordinate transformation, for the boundary-induced
energy-momentum tensor in coordinates $x^{i}$ one has (no summation over $l$%
):
\begin{eqnarray}
\langle T_{l}^{l}\rangle ^{\mathrm{(b)}} &=&[\xi /a(\eta )]^{D+1}\langle T_{%
\mathrm{(R)}l}^{l}\rangle ^{\mathrm{(b)}},\;l=0,3,\ldots ,D,  \notag \\
\langle T_{l}^{l}\rangle ^{\mathrm{(b)}} &=&[\xi /a(\eta )]^{D+1}\left[
\langle T_{\mathrm{(R)}l}^{l}\rangle ^{\mathrm{(b)}}+(-1)^{l}\Omega ^{2}\sin
^{2}\theta (\langle T_{\mathrm{(R)}1}^{1}\rangle ^{\mathrm{(b)}}-\langle T_{%
\mathrm{(R)}2}^{2}\rangle ^{\mathrm{(b)}})\right] ,\;l=1,2,
\label{Tik0dSstb} \\
\langle T_{1}^{2}\rangle ^{\mathrm{(b)}} &=&[\xi /a(\eta )]^{D+1}\Omega
^{2}\sin \theta \frac{\cos \theta -r\gamma }{r}(\langle T_{\mathrm{(R)}%
2}^{2}\rangle ^{\mathrm{(b)}}-\langle T_{\mathrm{(R)}1}^{1}\rangle ^{\mathrm{%
(b)}}),  \notag
\end{eqnarray}%
with the tensor $\langle T_{\mathrm{(R)}i}^{k}\rangle ^{\mathrm{(b)}}$ given
by (\ref{Tikb}). Similar to the case of the Rindler problem, the tensor $%
\langle T_{i}^{k}\rangle ^{\mathrm{(b)}}$ is decomposed into the single
boundary and the second boundary induced parts. The resulting
energy-momentum tensor is non-diagonal due to the anisotropy of the vacuum
stresses in the Rindler counterpart. The time dependence of the boundary
induced part in the VEV\ of the energy-momentum tensor appears in the form $%
a^{-D-1}(\eta )$. In addition this VEV\ depend on the radial coordinate $r$
and on the angle $\theta $. In the components $\langle T_{l}^{l}\rangle ^{%
\mathrm{(b)}}$ with $l=0,3,\ldots ,D$, the dependence on $r$ and $\theta $
enters in the combination $\xi =\xi (r,\theta )$. So, for a fixed $\eta $,
these components are constant on the lines $\xi =\mathrm{const}$ (see figure %
\ref{fig1}). This is not the case for the diagonal components $\langle
T_{l}^{l}\rangle ^{\mathrm{(b)}}$ with $l=1,2$, and for the off-diagonal
component. In the model with $D=3$ and with a power-law scale factor, $%
a(t)\sim t^{c}$, $t$ being the comoving time coordinate, the boundary-free
part (\ref{Tik0dSst}) behaves as $t^{-4}$. In this case, at early (late)
stages of the cosmological expansion the boundary induced part dominates for
$c>1$ ($c<1$).

\begin{figure}[tbph]
\begin{center}
\epsfig{figure=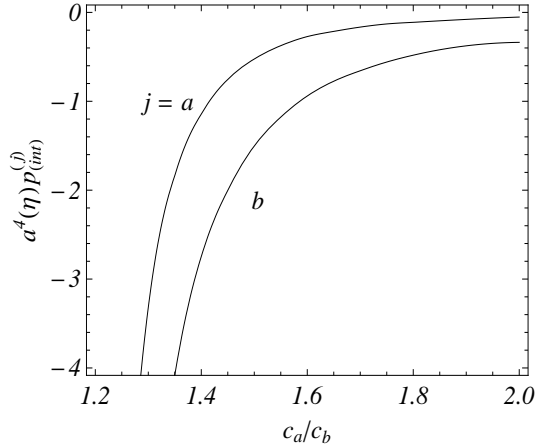,width=7.cm,height=6.cm}
\end{center}
\caption{Interaction parts of the vacuum pressure on the boundaries as
functions of the ratio $c_{a}/c_{b}$ for a scalar field with Neumann
boundary condition.}
\label{fig2}
\end{figure}

The $k$-th component of the Casimir force acting per unit surface of the
boundary at $\xi =\xi _{j}$ is determined by the expression $\langle
T_{l}^{k}\rangle ^{\mathrm{(b)}}n_{j}^{l}|_{\xi =\xi _{j}}$, where the
normal to the boundary is defined by relation (\ref{nl}). Using this
expression and formulae (\ref{Tik0dSstb}), it can be seen that the force is
presented in the form%
\begin{equation}
\langle T_{l}^{k}\rangle ^{\mathrm{(b)}}n_{j}^{l}|_{\xi =\xi
_{j}}=n_{j}^{k}[\xi
_{j}/a(\eta )]^{D+1}\langle T_{\mathrm{(R)}1}^{1}\rangle ^{\mathrm{(b)}%
}|_{\xi =\xi _{j}}.  \label{CasF1}
\end{equation}%
Hence, we conclude that the Casimir force is directed along the normal to
the boundary and does no depend on the point of the boundary. Note that the
quantity $p_{\mathrm{(R)}}^{(j)}=-\langle T_{\mathrm{(R)}1}^{1}\rangle ^{%
\mathrm{(b)}}|_{\xi =\xi _{j}}$ determines the effective pressure on the
plate at $\xi =\xi _{j}$ in the corresponding Rindler problem. As it was
shown in \cite{Avag02,Saha06}, the latter is presented in the form $%
p^{(j)}=p_{1}^{(j)}+p_{\mathrm{(R,int)}}^{(j)}$, where the first term on the
right is the pressure for a single plate at \ $\xi =\xi _{j}$ when the
second plate is absent and $p_{\mathrm{(R,int)}}^{(j)}$ is the pressure
induced by the presence of the second plate. The latter can be termed as the
interaction part of the effective pressure. The surface divergences are
contained in the single boundary parts (for a recent discussion of the
surface divergences see \cite{Milt10}) and the interaction parts are finite
for all non-zero distances between the boundaries. By using the expression
of $p_{\mathrm{(R,int)}}^{(j)}$, for the corresponding pressure in RW bulk
we find%
\begin{eqnarray}
p_{\mathrm{(int)}}^{(j)} &=&[\xi _{j}/a(\eta )]^{D+1}p_{\mathrm{(R,int)}%
}^{(j)}=\frac{2^{1-D}\pi ^{-(D+1)/2}}{\Gamma ((D-1)/2)a^{D+1}(\eta )}%
\int_{0}^{\infty }dx\,x^{D-2}\int_{0}^{\infty }d\omega \,  \notag \\
&&\times \left[ x^{2}+\omega ^{2}+(1-1/D)\delta _{j}\xi _{j}\beta _{\mathrm{R%
}j}-\xi _{j}{}^{2}\beta _{\mathrm{R}j}^{2}\right] \Omega _{j\omega }(x\xi
_{a}/\xi _{j},x\xi _{b}/\xi _{j}).  \label{pjint}
\end{eqnarray}%
In dependence of the coefficients in the boundary conditions the
corresponding forces can be either attractive or repulsive. In particular,
for Dirichlet boundary conditions ($\beta _{\mathrm{R}j}=\infty $) and for
Neumann boundary conditions in the Rindler spacetime problem ($\beta _{%
\mathrm{R}j}=0$) the interaction forces are always attractive. Note that the
latter corresponds to the Robin type problem in RW background (see (\ref%
{relcoef})). For Neumann boundary conditions in the RW problem we have $%
\delta _{j}\xi _{j}\beta _{\mathrm{R}j}=(D-1)/2$. In figure \ref{fig2}, the
interaction parts in the corresponding vacuum pressures are plotted as
functions of the ratio $c_{a}/c_{b}=\xi _{b}/\xi _{a}$. As it is seen from
the graphs, the forces are attractive in this case as well.

At small separations between the boundaries, corresponding to $%
c_{a}/c_{b}-1\ll 1$, for Dirichlet condition on one of boundaries and
non-Dirichlet boundary condition on the other the leading term in the
asymptotic expansion has the form%
\begin{equation}
p_{\mathrm{(int)}}^{(j)}\approx \frac{(1-2^{-D})D\zeta _{\text{R}}(D+1)}{%
(4\pi )^{(D+1)/2}a^{D+1}(\eta )}\Gamma ((D+1)/2),  \label{pjsmall}
\end{equation}%
with $\zeta _{\text{R}}(x)$ being the Riemann zeta function. The
corresponding forces are repulsive. For all other combinations of the
boundary conditions the leading term is given by%
\begin{equation}
p_{\mathrm{(int)}}^{(j)}\approx -\frac{D\zeta _{\text{R}}(D+1)}{(4\pi
)^{(D+1)/2}a^{D+1}(\eta )}\Gamma ((D+1)/2),  \label{pjsmall2}
\end{equation}%
and the forces are attractive. At large separations, $c_{a}/c_{b}\gg 1$, the
interaction parts decay as%
\begin{equation}
p_{\mathrm{(int)}}^{(a)}\propto \frac{(c_{b}/c_{a})^{D-1}}{a^{D+1}(\eta )\ln
^{3}(c_{a}/c_{b})},\;p_{\mathrm{(int)}}^{(b)}\propto \frac{1}{a^{D+1}(\eta
)\ln ^{2}(c_{a}/c_{b})}.  \label{plarge}
\end{equation}%
The nature of the corresponding forces depend on the coefficients in Robin
boundary conditions.

\section{Electromagnetic field}

\label{sec:Elmag}

In this section we consider the electromagnetic field in the region between
two boundaries described by (\ref{BoundEq}) with $j=a,b$. We assume that
these boundaries are perfect conductors with the boundary conditions of
vanishing of the normal component of the magnetic field and the tangential
components of the electric field, evaluated at the local inertial frame.
Since the electromagnetic field is conformal in $D=3$ we restrict ourselves
to this case. As in the previous section, we use the conformal relation
between the problems in RW and Rindler spacetimes. The VEV of the
energy-momentum tensor for the electromagnetic field in the region between
two conducting plates moving with uniform proper acceleration in the
Fulling-Rindler vacuum is investigated in Refs. \cite{Avag02}. Similar to
the scalar case, this expectation value is presented in the form (\ref{TikR}%
), where the boundary free part is given by the formula%
\begin{equation}
\langle \tilde{0}_{\mathrm{R}}|T_{i}^{k}[g_{lm}^{\mathrm{R}},\varphi _{%
\mathrm{R}}]|\tilde{0}_{\mathrm{R}}\rangle =\frac{11}{240\pi ^{2}\xi ^{4}}%
\mathrm{diag}\left( -1,1/3,1/3,1/3\right) .  \label{TikREl}
\end{equation}%
For the correction due to the boundaries we have the expression%
\begin{equation}
\langle T_{\mathrm{(R)}i}^{k}\rangle ^{\mathrm{(b)}}=\langle T_{\mathrm{(R)}%
i}^{k}\rangle ^{(j)}-\frac{\delta _{i}^{k}}{4\pi ^{2}}\int_{0}^{\infty
}dk\,k^{3}\int_{0}^{\infty }d\omega \,\sum_{\sigma =0,1}\Omega _{j\omega
}^{(\sigma )}(k\xi _{a},k\xi _{b})F_{\mathrm{em}}^{(i)}[Z_{\omega }^{(\sigma
)}(k\xi ,k\xi _{j})],  \label{TikbEln}
\end{equation}%
where the notations are defined as%
\begin{equation}
Z_{\omega }^{(\sigma )}(u,v)=\partial _{v}^{\sigma }\left[ I_{\omega
}(v)K_{\omega }(u)-K_{\omega }(v)I_{\omega }(u)\right] ,  \label{Zjel}
\end{equation}%
and%
\begin{equation}
\Omega _{a\omega }^{(\sigma )}(u,v)=\frac{K_{\omega }^{(\sigma
)}(v)/K_{\omega }^{(\sigma )}(u)}{\partial _{u}^{\sigma }Z_{\omega
}^{(\sigma )}(u,v)},\;\Omega _{b\omega }^{(\sigma )}(u,v)=\frac{I_{\omega
}^{(\sigma )}(u)/I_{\omega }^{(\sigma )}(v)}{\partial _{u}^{\sigma
}Z_{\omega }^{(\sigma )}(u,v)},  \label{Omjel}
\end{equation}%
For the functions $F_{\mathrm{em}}^{(i)}[g(z)]$ we have%
\begin{equation}
F_{\mathrm{em}}^{(i)}[g(z)]=(-1)^{i}g^{\prime 2}(z)+[1-(-1)^{i}\omega
^{2}/z^{2}]g^{2}(z),\;i=0,1,  \label{Fiem}
\end{equation}%
and $F_{\mathrm{em}}^{(i)}[g(z)]=-g^{2}(z)$ for $i=2,3$. In (\ref{Omjel}),
for the modified Bessel functions $f^{(0)}(u)=f(u)$ and $f^{(1)}(u)=f^{%
\prime }(u)$ with $f=I_{\omega },K_{\omega }$. In the region $\xi >\xi _{j}$
for the single boundary induced part one has the expression \cite%
{CandD,Saha02}
\begin{equation}
\langle T_{\mathrm{(R)}i}^{k}\rangle ^{(j)}=-\frac{\delta _{i}^{k}}{4\pi
^{2}\xi _{j}^{4}}\int_{0}^{\infty }dx\,x^{3}\int_{0}^{\infty }d\omega
\sum_{\sigma =0,1}\frac{I_{\omega }^{(\sigma )}(x)}{K_{\omega }^{(\sigma
)}(x)}F_{\mathrm{em}}^{(i)}[K_{\omega }(x\xi /\xi _{j})],  \label{TikbEl}
\end{equation}%
The corresponding formula in the region $\xi <\xi _{j}$ is obtained from (%
\ref{TikbEl}) by the replacements $I_{\omega }\rightleftarrows K_{\omega }$.
In formulae (\ref{TikbEln}) and (\ref{TikbEl}), the terms with $\sigma =0$
and $\sigma =1$ correspond to the contributions of the transverse electric
(TE) and transverse magnetic (TM) waves respectively.

The VEV of the energy-momentum tensor in the RW bulk is presented in the
form (\ref{TikdS1}), where the boundary free part is given by the expression%
\begin{equation}
\langle \tilde{0}_{\mathrm{RW}}|T_{i}^{k}\left[ g_{lm},\varphi \right] |%
\tilde{0}_{\mathrm{RW}}\rangle =\frac{11a^{-4}(\eta )}{240\pi ^{2}}\mathrm{%
diag}\left( -1,1/3,1/3,1/3\right) +\frac{62^{(3)}H_{i}^{k}+3^{(1)}H_{i}^{k}}{%
2880\pi ^{2}}.  \label{TikElfree}
\end{equation}%
The correction due to the boundaries is given by formulae (\ref{Tik0dSstb})
with the Rindler components from (\ref{TikbEln}). On the basis of (\ref%
{TikbEln}), these corrections are decomposed into single boundary and second
boundary induced parts.

\begin{figure}[tbph]
\begin{center}
\epsfig{figure=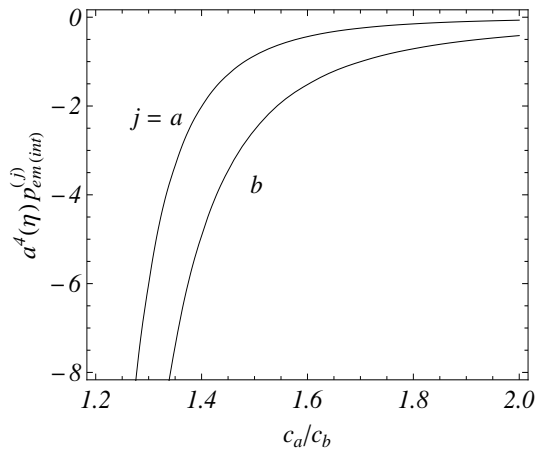,width=7.cm,height=6.cm}
\end{center}
\caption{The same as in figure \protect\ref{fig2} for the electromagnetic
field with perfect conductor conditions on the boundaries.}
\label{fig3}
\end{figure}

Now we turn to the investigation of the vacuum forces acting on boundaries.
As in the case of a scalar field, the electromagnetic Casimir force is
directed along the normal to the boundary. This force is decomposed into
self-action and interaction parts. The latter is induced by the presence of
the second boundary and is given by the expression%
\begin{equation}
p_{\mathrm{em(int)}}^{(j)}=-\frac{a^{-D-1}(\eta )}{4\pi ^{2}}%
\int_{0}^{\infty }dx\,x\int_{0}^{\infty }d\omega \,\sum_{\sigma
=0,1}(-1)^{\sigma }(1+\omega ^{2}/x^{2})^{\sigma }\Omega _{j\omega
}^{(\sigma )}(x\xi _{a}/\xi _{j},x\xi _{b}/\xi _{j}).  \label{pjel}
\end{equation}%
The corresponding forces are attractive. The interaction parts of the vacuum
pressures on the boundaries are plotted in figure \ref{fig3} as functions of
the ratio $c_{a}/c_{b}$. Note that $p_{\mathrm{em(int)}}^{(j)}$ is the sum
of the corresponding quantities for Dirichlet and Neumann (in the Rindler
spacetime problem) scalars and the asymptotics at small and large distances
directly follow from those for the scalar case.

\section{Conclusion}

\label{sec:Conc}

In the present paper we have considered an exactly solvable problem for the
Casimir effect with curved bulk and boundary geometries. In obtaining the
VEVs of the energy-momentum tensor for conformally coupled massless scalar
and electromagnetic fields, we use the conformal relation between the RW
spacetime with negative spatial curvature and Rindler spacetime. Boundaries
in the problem on RW background are described by (\ref{BoundEq}). They are
the conformal images of two parallel plates moving with constant proper
acceleration through the Fulling-Rindler vacuum. We assumed Robin boundary
conditions in the case of the scalar field and perfect conductor boundary
conditions for the electromagnetic field. For the corresponding Rindler
problem the Casimir densities are given by expressions (\ref{Tikb}) and (\ref%
{TikbEln}) for the scalar and electromagnetic fields respectively.

In order to generate the VEVs for the RW problem, first we use the
coordinate transformation (\ref{CordTrans}) which presents the RW metric in
the form manifestly conformal to the Rindler metric. As the next step, we
obtain the VEVs in new coordinates by the conformal transformation from the
results of the corresponding problem in Rindler spacetime. For the scalar
field the coefficients in Robin boundary conditions are related by (\ref%
{relcoef}). At the final stage, we transform the VEVs of the energy-momentum
tensor to the initial coordinates. In this way the VEVs are decomposed into
boundary free and boundary induced parts. The latter are given by
expressions (\ref{Tik0dSstb}). The vacuum stresses in the Rindler spacetime
problem are anisotropic and as a consequence of this the boundary induced
part in the vacuum energy-momentum tensor for the RW problem is non-diagonal.

Having the VEVs of the energy-momentum tensor we have investigated the
Casimir forces acting on the boundaries. These forces are decomposed into
self-action and interaction parts. The interaction forces are directed along
the normal to the boundary and the corresponding effective pressures are
given by expressions (\ref{pjint}) and (\ref{pjel}) for the scalar and
electromagnetic fields respectively. These pressures are independent of the
point on the boundary. For the scalar field, in dependence of the Robin
coefficients, the interaction forces can be either attractive or repulsive.
For Dirichlet and Neumann boundary conditions they are attractive for all
separations between the boundaries. In the case of the electromagnetic field
the force is the sum of the forces for Dirichlet and Neumann scalars and is
attractive.

\section*{Acknowledgments}

A.A.S. was supported by Conselho Nacional de Desenvolvimento Cient\'{\i}fico
e Tecnol\'{o}gico (CNPq) and by the Armenian Ministry of Education and
Science Grant No. 119. The work of M.R.S. has been supported by Research
Institute for Astronomy and Astrophysics of Maragha, Iran.

\end{document}